\documentclass[prl,showpacs,showkeys,nofootinbib,twocolumn,floatfix,superscriptaddress]{revtex4}

\usepackage{multirow}
\usepackage{graphicx}  %
\usepackage{amsmath}
\usepackage{amsfonts}
\usepackage{ulem}
\usepackage{bm}  %
\usepackage{xcolor}
\usepackage{hyperref}
\hypersetup{
    colorlinks,
    linkcolor={red!50!black},
    citecolor={blue!50!black},
    urlcolor={blue!80!black}
}

\newcommand{\bea}{\begin{eqnarray}}
\newcommand{\eea}{\end{eqnarray}}
\newcommand{\beq}{\begin{equation}}
\newcommand{\eeq}{\end{equation}}
\newcommand{\bqa}{\begin{eqnarray}}
\newcommand{\eqa}{\end{eqnarray}}

\def\mqo2{{\!\!\!}}

\begin{document}

\title{
Universal Relations for Range Corrections to Efimov Features
}
\author{Chen Ji}
\affiliation{TRIUMF, 4004 Wesbrook Mall, Vancouver,
British Columbia V6T 2A3, Canada}
\email{jichen@triumf.ca}
\author{Eric Braaten}
\affiliation{Department of Physics, The Ohio State University, Columbus, Ohio 43210, USA}
\email{braaten.1@osu.edu}
\author{Daniel R. Phillips}
\affiliation{Institute of Nuclear and Particle Physics and Department of Physics and Astronomy, Ohio University, Athens, Ohio 45701, USA}
\email{phillid1@ohio.edu}
\author{Lucas Platter}
\affiliation{Physics Division, Argonne National Laboratory, Argonne, Illinois 60439, USA}
\affiliation{Department of Physics and Astronomy, University of Tennessee, Knoxville, TN 37996, USA}
\affiliation{Physics Division, Oak Ridge National Laboratory, Oak Ridge, TN 37831, USA}
\email{lplatter@utk.edu}
\date{\today}

\begin{abstract}
  In a three-body system of identical bosons interacting through a
  large S-wave scattering length $a$, there are several sets of
  Efimov features related by discrete scale invariance. Effective
  field theory was recently used to derive universal relations
  between these Efimov features that include the first-order
  correction due to a nonzero effective range $r_s$.  We reveal a
  simple pattern in these range corrections that had not been
  previously identified. The pattern is explained by the
  renormalization group for the 
  effective field theory, which implies that the Efimov
  three-body parameter runs logarithmically with the momentum scale at
  a rate proportional to $r_s/a$.  The running Efimov parameter also
  explains the empirical observation that range corrections can be
  largely taken into account by shifting the Efimov parameter by an
  adjustable parameter divided by $a$.  The accuracy of universal
  relations that include first-order range corrections is verified by
  comparing with various theoretical calculations using models with
  nonzero range.
\end{abstract}
\smallskip
\maketitle

{\it Introduction.}
Dramatic experimental and theoretical progress in few-body physics has
been stimulated by the realization that Feshbach resonances can be
used to control the strength of interatomic interactions in ultracold
atomic gases.  The {\it low-energy universality} that arises when
scattering lengths are much larger than the range of interactions is
of particularly broad interest.  It implies that for systems as
disparate as atoms, hadrons, and nuclei, dimensionless combinations of
few-body observables are the same, despite orders of magnitude
differences in the length and energy scales.  

A particularly fascinating class of low-energy universal behavior is 
{\it Efimov  physics}, which is characterized by {\it discrete scale invariance}
\cite{Braaten:2004rn,Hammer:2010kp}.  The ability to control the strength of
interatomic interactions has allowed the observations of various
aspects of Efimov physics in ultracold atoms \cite{FZBGHNG:1108}.  The
simplest example of Efimov physics is the {\it Efimov effect}: in the
unitary limit where the scattering length $a$ is infinite, there are
infinitely many three-body bound states with an accumulation point at
the 3-particle scattering threshold \cite{Efimov70}.  These Efimov trimers have
binding energies $E_{T,n}$ whose limiting behavior is
\begin{equation}
\label{eq:efimov-spectrum}
E_{T,n} \longrightarrow
  \lambda^{-2n}\hbar^2 \kappa_*^2/m~~~{\rm as}~n \to \infty,
\end{equation}
where $m$ is the particle mass and $\lambda$ is called the {\it
  discrete scaling factor}.  The limit in
Eq.~\eqref{eq:efimov-spectrum} defines a three-body interaction
parameter $\kappa_*$.  In the case of identical bosons,
$\lambda=e^{\pi/s_0} = 22.6944$ with $s_0=1.00624$.  The discrete
spectrum in the unitary limit in Eq.~\eqref{eq:efimov-spectrum}
reveals that low-energy physics in the three-particle sector depends
not only on the scattering length $a$ but also on $\kappa_*$.  Efimov
physics can also be revealed through discrete scale invariance at
finite values of $a$.  For example, the negative scattering lengths $a_{-,n}$
at which the Efimov trimers cross the three-boson threshold have the
limiting behavior
\begin{equation}
\label{eq:unia-}
a_{-,n} \longrightarrow  \theta_- \,  \lambda^{n}  \kappa_*^{-1}
~~~{\rm as}~n \to \infty,
\end{equation}
where $\theta_-=-1.50763$~\cite{GME:0802} is a universal number.

The discrete scale invariance that characterizes Efimov physics
becomes exact, with the limits in Eqs.~\eqref{eq:efimov-spectrum} and
\eqref{eq:unia-} replaced by equalities, only in the limit of
zero-range interactions.  In that limit, few-body observables depend
only on $a$ and $\kappa_*$.  The interactions in real physical systems
always have nonzero range.  Experiments naturally involve the deepest
Efimov trimers, for which range corrections are largest.  In order to
make quantitative experimental tests of Efimov physics, it is
important to understand the range corrections in detail.

Range corrections can be studied theoretically by calculating Efimov
features in various models for interatomic interactions and comparing
them with the universal zero-range predictions.  Kievsky and
Gattobigio discovered empirically that range corrections can be
largely taken into account by making substitutions for $a$ and
$\kappa_*$ in zero-range formulas \cite{Kievsky:2012ss}.  Their
substitution for the Efimov parameter is
$\kappa_* \to \kappa_* + \Gamma/a$, where $\Gamma$ is determined
empirically for each observable and each system.

A systematic theoretical approach to the problem is to organize range
corrections into an expansion in powers of the range $r_0$~\cite{Hammer:2001gh}.  In the
two-body sector, the only new parameter that enters through second
order in $r_0$ is the S-wave effective range $r_s$. Bedaque et al.\
showed that three-body observables do not depend on any additional
three-body parameter at first order in $r_0$ \cite{Bedaque:2002yg}, but
they made the implicit assumption that the scattering length $a$ is
fixed. Ji, Phillips, and Platter (JPP) showed that if variations in
the scattering length are considered, there is an additional
three-body parameter at first order in $r_0$ \cite{Ji:2010su}. They
developed an effective field theory (EFT) framework for calculating
range corrections \cite{Ji:2011qg,Ji:2010su}, and used it to derive
universal relations between Efimov features, such as $E_{T,n}$ and
$a_{-,n}$, that are accurate to next-to-leading order (NLO) in
$r_0/a$.

In this paper, we reveal a pattern in the first-order range
corrections to Efimov features that was not recognized in
Refs.~\cite{Ji:2011qg,Ji:2010su}.  We demonstrate that the pattern has
a simple renormalization-group  interpretation in terms of a 
{\it  running Efimov parameter} that runs with the momentum scale at a
rate proportional to $r_s/a$.  The empirical shift of the Efimov
parameter used in Ref.~\cite{Kievsky:2012ss} to incorporate range
corrections into zero-range formulas can be identified as the
expansion of the running Efimov parameter to first order in $r_s/a$.
We compare the predictions of the NLO universal relations with results
for Efimov features in specific models.

{\it Efimov Features.}
The Efimov trimers can be labelled by an integer $n$. 
In the 3-atom sector, the most dramatic Efimov features 
associated with the $n^{\rm th}$ branch of Efimov trimers are
(a) the binding momentum $\kappa_{T,n} = (mE_{T,n}/\hbar^2)^{1/2}$ 
of the Efimov trimer in the unitary limit $a = \pm \infty$,
(b) the negative scattering length  $a_{-,n}$
at which the Efimov trimer crosses the 3-atom threshold,
(c) the positive scattering length  $a_{*,n}$
at which the Efimov trimer disappears through the atom-dimer threshold,
and (d) the positive scattering length  $a_{+,n}$
at which there is an interference minimum 
in the three-body recombination rate at threshold.

The three-body interaction parameter $\kappa_*$ defined by
Eq.~\eqref{eq:efimov-spectrum} is approximately equal to the binding
momentum of the Efimov trimer labelled $n=0$.  The binding momenta are
$\kappa_{T,n} = \lambda^{-n} \kappa_*$ in the zero-range limit.  JPP
showed that this equation remains exact, with no range corrections,
at first order in $r_0/a$ \cite{Platter:2008cx}. In the
universal zero-range limit, the ratio of any pair of Efimov features
is a universal number.  The leading order (LO) universal relations 
are
\begin{equation}
\label{eq:a-kappa*}
a_{i,n} = \lambda^{n} \theta_i \kappa_*^{-1},
\end{equation}
where $\theta_-=-1.50763$,
$\theta_+= |\theta_-|/\sqrt{\lambda}=0.316473$~\cite{GME:0802}, and
$\theta_* = 0.0707645$~\cite{Braaten:2004rn}.

{\it First-order Range Corrections.}  JPP developed an EFT framework
for calculating range corrections as strict expansions in powers of
$r_0$ \cite{Ji:2010su,Ji:2011qg}.  At first order in $r_0$, two Efimov
features are required as inputs, with at most one being a trimer
binding momentum (or $\kappa_*$).  A simple choice for the two Efimov
features is $\kappa_*$ and $a_{-,0}$. The deviation
from the zero-range prediction for $a_{-,0}$ can be expressed using
Eq.~\eqref{eq:a-kappa*} as
$1/a_{-,0} = \theta_{-}^{-1} \kappa_* + {\cal I} \kappa_*^2 r_s$,
which defines a nonuniversal number ${\cal I}$.  JPP showed that the
range expansion to first order in the range for any other Efimov
feature can then be expressed as
\begin{equation}
\label{eq:a--kappa*JPP}
1/a_{i,n} = \lambda^{-n} \theta_i^{-1} \kappa_*
 +  \left( \xi_{i,n}  + \eta_{i,n}  {\cal I} \right) \kappa_*^2 r_s~,
\end{equation}
where $\xi_{i,n}$ and $\eta_{i,n}$ are universal numbers. JPP
calculated many such numbers to at least three digits \cite{Ji:2010su,Ji:2011qg}.  

There is a pattern to
the dependence of the universal numbers in Eq.~\eqref{eq:a--kappa*JPP}
on the number $n$ labelling the branch of Efimov trimers that was not
identified in Ref.~\cite{Ji:2011qg}.  The range expansion in
Eq.~\eqref{eq:a--kappa*JPP} can be expressed in the much simpler form
\begin{equation}
\label{eq:a--kappa*NLO}
a_{i,n} = \lambda^{n} \theta_i \kappa_*^{-1}
+ (J_i - n \sigma)  r_s,
\end{equation}
where $\sigma = 1.095$ is a universal number. The differences between
the coefficients $J_i$ are also universal numbers:
$J_+ - J_- = 0.548$, $J_* - J_- = 1.250$.  We will refer to
Eq.~\eqref{eq:a--kappa*NLO} as the {\it NLO range expansion} for the
Efimov feature.

{\it Renormalization}.  The $n$ in the range correction in
Eq.~\eqref{eq:a--kappa*NLO} can be understood as a logarithmic dependence 
on the scale at which the observable $a_{i,n}$ is measured,
suggesting a
renormalization-group interpretation.  We therefore discuss the
renormalization of the EFT used to calculate range corrections in
Refs.~\cite{Ji:2010su,Ji:2011qg}.  The Lagrangian density for that
EFT, which has an atom field $\psi$ and a molecule field $\phi$, is
\begin{eqnarray}
\nonumber
\mathcal{L}&=&
\psi^\dagger \left(i\partial_0 +  \frac{\nabla^2}{2m}\right) \psi 
- \phi^\dagger \left(i\partial_0 +  \frac{\nabla^2}{4m}-\Delta \right) \phi
\\
&& -\frac{g}{\sqrt{2}}\left(\phi^\dagger\psi\psi+\textrm{h.c}\right)
+h\phi^\dagger \phi \psi^\dagger \psi.
\label{eq:Lagrangian}
\end{eqnarray}
The parameters $g$ and $\Delta$ can be tuned as functions of the
ultraviolet cutoff, $\Lambda$, so that $a$ and $r_s$ have the desired
values. In the three-body sector, the ultraviolet cutoff can be
implemented as an upper limit $\Lambda$ on the loop momentum in a
modified Skorniakov--Ter-Martirosian equation \cite{STM57} for the scattering of
three bosons~\cite{Bedaque:1998km}.
The three-body coupling constant can be expressed as
$h = 2mg^2H/\Lambda^2$, where $H$ is a dimensionless log-periodic
function of $\Lambda$.  Its dependence on $\Lambda$ in the zero-range
limit was determined up to a numerical factor in
Ref.~\cite{Bedaque:1998km}:
\begin{equation}
\label{eq:H0}
H_0(\Lambda/\kappa_*)=0.879\, 
  \frac{\sin[s_0\log(\Lambda/\Lambda_*)+\arctan s_0]}
{\sin[s_0\log(\Lambda/\Lambda_*)-\arctan s_0]}~,
\end{equation}
where $\Lambda_* = 0.548\, \kappa_*$. The multiplicative constant 
in Eq.~\eqref{eq:H0} was
first determined in Ref.~\cite{Braaten:2011sz}.  The dependence of $H$
on $\Lambda$ at first order in the range has the form~\cite{Ji:2011qg}
\begin{eqnarray}
\label{eq:H-expansion}
H(\Lambda) &=& H_0(\Lambda/\kappa_*) 
+  h_{10}(\Lambda/\kappa_*) \Lambda r_s
\nonumber
\\
&+& \left[ \gamma \, H_0'(\Lambda/\kappa_*)
\log(\Lambda/\mu_0)+\tilde{h}_{11}(\Lambda/\kappa_*) \right]  \frac{r_s}{a},
\end{eqnarray}
where $\mu_0$ is a momentum scale and $H_0'= (\Lambda d/d\Lambda) H_0$
is the logarithmic derivative of $H_0$ in Eq.~\eqref{eq:H0}. We will
not need the analytic forms of the log-periodic functions $h_{10}$ and
$\tilde{h}_{11}$. 
JPP gave an analytic expression for
$\gamma$ which is within 5\% of the value we find numerically,
$\gamma=0.351$.  In Ref.~\cite{Ji:2011qg}, $\mu_0$ was chosen so
that a second Efimov feature in addition to $\kappa_*$, such as
$a_{-,0}$, had no first-order range corrections.  The order of
magnitude of the required value of $\mu_0$ is then $1/|a_{-,0}|$.

{\it Running Efimov Parameter.}  The term proportional to
$\log(\Lambda/\mu_0)$ in Eq.~\eqref{eq:H-expansion} represents a
logarithmic violation of discrete scale invariance.  It can be
absorbed into the zero-range coupling constant by changing the
argument of $H_0$ to
$(\Lambda/\kappa_*)(\Lambda/\mu_0)^{\gamma r_s/a}$.  The power of
$\mu_0$ in the argument can be canceled by replacing $\kappa_*$ by a
{\it running Efimov parameter} 
defined by
\begin{equation}
  \label{eq:runningkappastar}
 \bar{\kappa}_*(\mu_0,a) \equiv
  (\mu_0/\kappa_*)^{-\gamma r_s/a} \,  \kappa_*~.
\end{equation}
Unnecessarily large logarithms in the range corrections to an
observable dominated by the momentum scale $Q$ can  be avoided by
expressing the observable in terms of $\bar{\kappa}_*(Q,a)$, $a$, and
$r_s$ instead of $\kappa_*$, $a$, and $r_s$.

In the NLO range expansion in Eq.~\eqref{eq:a--kappa*NLO}, the term
linear in $n$ gives unnecessarily large range corrections if $n$ is
large.  Natural scaling variables in the zero-range limit are the
inverse scattering length $1/a$ and the energy variable
$K = {\rm sign}(E)(m|E|/\hbar^2)^{1/2}$.  The momentum scale $Q$ for
an observable with energy $E$ at scattering length $a$ is
$Q=(K^2 + 1/a^2)^{1/2}$.  The momentum scales for the Efimov features
$a_{-,n}$, $a_{+,n}$, and $a_{*,n}$ are $1/|a_{-,n}|$, $1/a_{+,n}$,
and $\sqrt{2}/a_{*,n}$, respectively.  Since a factor of $\sqrt{2}$ is
not essential, these can be summarized via the LO universal relation
as $Q \approx \lambda^{-n} \kappa_*/|\theta_i|$.  If we replace
$\kappa_*$ in the LO universal relation in Eq.~\eqref{eq:a-kappa*} by
$\bar{\kappa}_*( Q,a)$ where $a = \lambda^{n} \theta_i \kappa_*^{-1}$
and expand in powers of $r_s$, we obtain a term
$-n \gamma r_s \log \lambda$.  This matches the $-n \sigma r_s$ term
in Eq.~\eqref{eq:a--kappa*NLO} provided
$\sigma = \gamma \log \lambda$.  Our numerically determined values of
$\gamma$ and $\sigma$ satisfy this condition to within numerical
accuracy.  Replacing $\kappa_*$ in the LO universal relation by
$\bar{\kappa}_*( \lambda^{-n} \kappa_*/|\theta_i|,\lambda^{n} \theta_i \kappa_*^{-1})$ 
therefore includes the NLO correction
proportional to $n r_s$.

Similarly, renormalization-group improvement of the NLO range expansion can be
obtained by eliminating $\kappa_*$ in Eq.~\eqref{eq:a--kappa*NLO} in
favor of the appropriate running Efimov parameter and demanding
agreement to first order in $r_s$:
\begin{equation}
\label{eq:a--kappa*RG}
a_{i,n} = 
\lambda^n \theta_i \left( \lambda^{n} |\theta_i|  \right)^{- \gamma r_s \kappa_*/(\lambda^{n} \theta_i)}
\kappa_*^{-1}
+ \tilde J_i   r_s,
\end{equation}
where $\tilde{J}_i=J_i + \gamma \log |\theta_i|$. The differences between
the coefficients $\tilde J_i$ are universal numbers:
$\tilde J_+ - \tilde J_- = 0.000$,  $\tilde J_* - \tilde J_- = 0.177$.
We will refer to Eq.~\eqref{eq:a--kappa*RG} as the {\it RG-improved
  NLO range expansion} for the Efimov feature.  For $n=1$, the NLO
range expansion in Eq.~\eqref{eq:a--kappa*NLO} and the RG-improved NLO
range expansion in Eq.~\eqref{eq:a--kappa*RG} have the same parametric
accuracy: higher-order range corrections are suppressed by a factor of
$ (\kappa_* r_s)^2$.  For large $n$, the NLO range expansion has
corrections of order $n^2 (\lambda^{-n}\kappa_* r_s)^2$.  In the
RG-improved NLO range expansion, those higher-order range corrections
that are enhanced by a factor of $n$ for every factor of $r_s$ are
summed up to all orders. Thus the higher-order range corrections to
Eq.~\eqref{eq:a--kappa*RG} are at most of order
$n (\lambda^{-n}\kappa_* r_s)^2$.

{\it Comparisons with Models.}  Successive Efimov features have been
calculated in several models with nonzero range. We can use the
results to illustrate the accuracy of the NLO range expansion in
Eq.~\eqref{eq:a--kappa*NLO} and the RG-improved NLO range expansion in
Eq.~\eqref{eq:a--kappa*RG}.  The NLO range expansion can be used to
express $a_{i,n+1}/(\lambda a_{n,i})$ as a linear combination of
$\kappa_*r_s$ and $J_i \kappa_*r_s$ with universal coefficients.  For
any three ratios of successive Efimov features, there is a linear
combination with universal coefficients in which $\kappa_*r_s$ is
eliminated.  We refer to this equation as an {\it NLO universal
  relation}.  If two of the ratios are taken as inputs, the third can
be predicted to NLO accuracy without knowing $r_s$.  Similarly the
RG-improved NLO range expansion can be used to derive a universal
relation that expresses $\log(a_{i,n+1}/(\lambda a_{i,n}))$ as a
linear combination of two other such expressions with universal
coefficients.

\begin{table}[tbh]
  \caption{\label{tbl:deltuva2}
    Ratios of successive Efimov features $a_{-,n+1}/a_{-,n}$
    divided by the discrete scaling factor  $\lambda$.
    The ratios of the features calculated by Deltuva \cite{Deltuva:2012zy}
    are compared to the predictions of  
    Eq.~\eqref{eq:a--kappa*NLO} (NLO)
    and  Eq.~\eqref{eq:a--kappa*RG} (RG-NLO)
    using the numbers in square brackets as inputs.}
\begin{center}
\begin{tabular}{l|ccccc|}
& \multicolumn{5}{c|}{$(a_{-,n+1}/a_{-,n})/\lambda$}  \\
\hline
~~~~~$n$  &  $0$  &  1  &  2  &  3  &  4    \\
\hline
Ref.~\cite{Deltuva:2012zy}~   & 0.7822 & 0.9665 & 0.9976 & 0.9999 & 1.0000 \\
NLO~  & 
[{\it 0.7822}] & [{\it 0.9665}] & 0.9975 & 0.9998 & 1.0000 \\
RG-NLO~  & 
[{\it 0.7822}] & [{\it 0.9665}] & 0.9975 & 0.9998 & 1.0000 \\
\hline
\end{tabular}
\end{center}
\end{table}

In Ref.~\cite{Deltuva:2012zy}, Deltuva calculated the scattering lengths 
at which universal tetramers cross the 4-boson threshold
for identical bosons interacting through a separable Gaussian potential.
He also gave accurate results for the ratios
$a_{-,n+1}/a_{-,n}$ of the scattering lengths at which 6 successive
Efimov trimers cross the three-boson threshold.
These ratios divided by $\lambda$ are given
in Table~\ref{tbl:deltuva2}.  They rapidly approach 1 as $n$
increases.  In Table~\ref{tbl:deltuva2}, we have taken the ratios
$a_{-,n+1}/ a_{-,n}$ for $n=0$ and 1 as inputs, and then used 
NLO universal relations to predict the ratios for $n=2$, $3$, and $4$.
The predictions are in excellent agreement with the results calculated
by Deltuva. The RG-improved NLO universal relation gives 
the same predictions to four digits.

\begin{table}[ht]
  \caption{\label{tbl:srz2}
    Ratios of successive Efimov features $a_{i,n+1}/ a_{i,n}$
    divided by the discrete scaling factor $\lambda$.
    The ratios calculated by Schmidt et al.~\cite{Schmidt:2012yn}
    are compared to the predictions of Eq.~\eqref{eq:a--kappa*NLO} (NLO)
    and Eq.~\eqref{eq:a--kappa*RG} (RG-NLO)
    using the numbers in square brackets as inputs.
    The three blocks are for models with $s_{\rm res}=100$, 1, and 0.1.
  }
\begin{center}
\begin{tabular}{l|ccc|ccc|}
& \multicolumn{3}{c|}{$(a_{-,n+1}/ a_{-,n})/\lambda$} 
& \multicolumn{3}{c|}{$(a_{*,n+1}/a_{*,n})/\lambda$} \\
\hline
~~~~~$n$   & 0  & 1  & 2 &  0  & 1  & 2     \\
\hline 
Ref.~\cite{Schmidt:2012yn}  & 
0.753 & 0.962 & 0.998 & 0.175 & 1.764 & 1.029  \\
NLO  & 
[{\it 0.753}] & [{\it 0.962}] & 0.997 & $-$8.814~~ & 1.150 & 1.032  \\
RG-NLO   & 
[{\it 0.753}] & [{\it 0.962}] & 0.997 & ~~0.0002 & 1.206 & 1.034   \\
\hline
Ref.~\cite{Schmidt:2012yn}  & 
1.008 & 0.998 & 0.9998 & 0.757 & 0.983 & 1.001  \\
NLO  & 
[{\it 1.008}] & [{\it 0.998}] & 0.9998 & $-$0.431~~ & 0.986 & 1.002  \\
RG-NLO  & 
[{\it 1.008}] & [{\it 0.998}] & 0.9998 & 0.240 & 0.986 & 1.002   \\
\hline
Ref.~\cite{Schmidt:2012yn}  & 
1.156 & 1.012 & 1.0007 & 1.188 & 0.938 & 0.991  \\
NLO & 
[{\it 1.156}] & [{\it 1.012}] & 1.0007 & 0.449 & 0.869 & 0.990   \\
RG-NLO  & 
[{\it 1.156}] & [{\it 1.012}] & 1.0008 & 0.916 & 0.887 & 0.990   \\
\hline
\end{tabular}
\end{center}
\end{table}

Schmidt et al.\ calculated multiple Efimov features for a
three-parameter model in which the interaction is a transition between an
atom pair and a molecule with a Gaussian form factor
\cite{Schmidt:2012yn}.  The three independent parameters can be
specified by $a$, $r_s$, and a dimensionless parameter
$s_{\rm res}=r_0/ r_*$  obtained by dividing the range $r_0$
of the Gaussian form factor by a length $r_*$ determined by its strength.
They presented results for three sets of parameters with $s_{\rm res}=100$,
1, and 0.1.  Their results for $a_{-,n+1}/a_{-,n}$ and
$a_{*,n+1}/a_{*,n}$ divided by the discrete scaling factor $\lambda$
are given in Table~\ref{tbl:srz2}.  We have taken the ratios
$a_{-,1}/ a_{-,0}$ and $a_{-,2}/ a_{-,1}$ as inputs, and then used NLO
universal relations to predict the other ratios.  All predictions for
$n=2$ are in good agreement with the results of
Ref.~\cite{Schmidt:2012yn}.  The predictions for $n=1$ are also in
reasonable agreement, except for the case $s_{\rm res}=100$.  The
large error in this case can be attributed to the feature $a_{-,0}$
being outside the window of universality.  The RG-improved NLO
universal relation gives slightly better predictions for
$n=1$ and dramatically better predictions for $a_{*,1}/a_{*,0}$,
for which
the NLO universal relation yields unphysical negative values 
in the $s_{\rm res}=100$ and 1 models.

{\it Shift in Three-body Parameter.}  In Ref.~\cite{Kievsky:2012ss},
Kievsky and Gattobigio calculated the binding energies of Efimov
trimers and the atom-dimer scattering length in two models with a
Gaussian two-body potential and with or without a Gaussian three-body
potential.  They made the empirical observation that range corrections
could be largely taken into account by making substitutions in the
zero-range formulas, which are functions of $a$ and $\kappa_*$ only:
(a) replace $a$ by the inverse binding momentum of the universal dimer
or virtual state, (b) replace the Efimov parameter $\kappa_*$ by
$\kappa_* + \Gamma/a$, where $\Gamma$ is a parameter that is
determined empirically for each observable and each system.  In
Ref.~\cite{Garrido:2013uxa}, Garrido et al.\ showed that this
prescription also works for the three-body recombination rate at
threshold.  The accuracy of the prescription was verified only in
limited regions of $a$ and for a few specific models.  The running
Efimov parameter provides a theoretical justification for the
prescription.  The expansion of the running Efimov parameter in
Eq.~\eqref{eq:runningkappastar} to first order in the range is
\begin{equation}
  \label{eq:runningkappastar1}
  \bar{\kappa}_*(Q,a) \approx
  \kappa_* \left[ 1 -
\gamma \log(Q/\kappa_*) r_s/a \right].
\end{equation}
If the slow logarithmic dependence of the momentum scale $Q$ on $a$ is
ignored, this has the same form $\kappa_* + \Gamma/a$ as the empirical
shift in the Efimov parameter introduced by Kievsky and Gattobigio.

The NLO range expansion can be used to identify universal relations
between the empirical constants $\Gamma$ for different 3-body
observables.  The prescription in Ref.~\cite{Garrido:2013uxa} for Efimov
features associated with the second Efimov trimer is
$a_{i,1} \kappa_{T,1} + \Gamma_{i}=\theta_{i}$.
According to the
NLO range expansion in Eq.~\eqref{eq:a--kappa*NLO}, the ratio
$(\Gamma_i-\Gamma_j)/(\kappa_{T,1} r_s)$ should be the universal
number $J_j - J_i$. The predicted universal value of
$ J_{*} - J_{+} $ is 0.702, while the two models  
considered in Ref.~\cite{Garrido:2013uxa} give the  results
0.671 and 0.753. We consider this to be reasonable agreement.
The corresponding ratios involving $\Gamma_-$ 
display large discrepancies between the two models, and do 
not agree as well with the universal predictions. 

{\it Experiment}.  By measuring the first Efimov feature $a_-^{(0)}$
at different Feshbach resonances in $^{133}$Cs atoms and comparing
with previous measurements with other atoms, the Innsbruck group
discovered a correlation to within 20\% between $a_-^{(0)}$ and the
coefficient of the van der Waals tail $-C_6/r^6$ of the interatomic
potential: $a_{-,0}\approx -9 \, r_{\rm vdW}$, where
$r_{\rm vdW} = \frac12(mC_6/\hbar^2)^{1/4}$ \cite{Innsbruch-1106}.
This correlation, which can be called {\it van der Waals
  universality}, was subsequently verified
theoretically~\cite{WDEG-1201,NEU-1208}.  Van der Waals universality
gives a sharp prediction for the effective range: $r_s=2.79r_{\rm vdW}$ \cite{BoGao:1998}. It thus predicts
a narrow range of values for the coefficient $J_-$ in
Eq.~\eqref{eq:a--kappa*NLO}.

To use the NLO range expansion to predict Efimov features for a
specific bosonic atom, the required inputs are the scattering length
$a$, the effective range $r_s$, and two  measured Efimov
features.  There are not many atoms for which there are accurate
measurements of three or more Efimov features.  Several Efimov
features have been observed for both $^7$Li and ${}^{133}$Cs, but
there are complications in $^7$Li from significant variation of $r_s$
and in ${}^{133}$Cs from multiple Feshbach resonances.  Comparison of
NLO predictions with experimentally measured Efimov features will be
presented elsewhere \cite{Ji:2015?}.

{\it Summary}. We have shown that universal relations which account for
a nonzero effective range are able to predict Efimov features with
higher accuracy.  A simple pattern in the NLO range expansion can be
interpreted in terms of a running Efimov parameter.  The running
Efimov parameter also explains the empirical findings of Gattobigio
et al.\ that range corrections can be incorporated into universal
zero-range results through a shift in the Efimov three-body parameter.

~

\acknowledgments {This work was supported by the Office of Nuclear
  Physics, U.S.~Department of Energy under Contract
  nos.\ DE-AC02-06CH11357, DE-AC05-00OR22725, and DE-FG02-93ER40756; by the
  National Science Foundation under grant PHY-131086; by the Natural
  Sciences and Engineering Research Council (NSERC) and the National
  Research Council of Canada; and by the Simons Foundation}


\end{document}